# Mundus vult decipi, ergo decipiatur: Visual Communication of Uncertainty in Election Polls

Alexander Bauer[1] · André Klima[1] · Jana Gauß[1] · Hannah Kümpel[1] · Andreas Bender[1] · Helmut Küchenhoff[1]

**Abstract**

Election poll reporting often focuses on mean values and only subordinately discusses the underlying uncertainty. Subsequent interpretations are too often phrased as certain. Moreover, media coverage rarely adequately takes into account the differences between now- and forecasts. These challenges were ubiquitous in the context of the 2016 and 2020 U.S. presidential elections, but are also present in multi-party systems like Germany. We discuss potential sources of bias in nowcasting and forecasting and review the current standards in the visual presentation of survey-based nowcasts. Concepts are presented to attenuate the issue of falsely perceived accuracy. We discuss multiple visual presentation techniques for central aspects in poll reporting. One key idea is the use of Probabilities of Events instead of party shares. The presented ideas offer modern and improved ways to communicate (changes in) the electoral mood for the general media.

---

[1] Statistical Consulting Unit StaBLab, Department of Statistics, LMU München, Munich, Germany

# 1 Introduction

Since their beginnings in the 1930s, election polls have always remained a controversial topic. The inherent difficulties evaluating the adequacy of survey designs combined with the media's tendency to oversimplify can lead to highly problematic reporting of election polls; especially since polls published shortly before election day have been shown to influence the voting behavior (Moy & Rinke, 2012).

A recent example of controversial election reporting was the 2016 U.S. presidential election. Media coverage shortly before election day focused on Donald Trump having only a small chance of winning the election, with most analysts reporting far lower probabilities of a Trump victory than the 28.6% reported by FiveThirtyEight (FiveThirtyEight 2016). In the aftermath of the election, pollsters and their methods were not only scrutinized but often bluntly attacked (e.g. Siegel, 2016). Besides substantial methodological shortcomings (e.g. with covering all subpopulations) a major issue was that even the 71.4% chance of Hillary Clinton winning was widely perceived as her victory being (virtually) certain. This can, to a large extent, be attributed to inadequate communication of the uncertainty behind opinion polls and forecasts, as well as to general statistical (il)literacy.

Overall, election poll reporting in (mass) media is inherently shaped by a seemingly unsolvable problem: Communicate the observed situation in simple terms, while keeping the consumer aware of the underlying (un)certainty in its entire complexity. In this article, we will showcase the most common pitfalls when communicating election poll results, and show alternative ways of how to approach them in a statistically adequate way. Selected visualizations are presented on koala.stat.uni-muenchen.de for German election polls.

In Chapter 2 we give a brief introduction to the basic concepts of election poll reporting and potential sources of bias. Chapter 3 shows current examples of how election polls get reported. Alternatives are proposed in Chapter 4 and discussed in Chapter 5.

# 2 Election Poll Reporting - Uncertainties and Biases

An ideal election poll would represent a simple random sample among all eligible voters. Every person would have the same probability of being sampled, and each selected person would participate and answer truthfully. In practice, this is never the case and different strategies are used for handling non-response and incorrect answers. Different survey designs exist for opinion polling, the most prominent ones being interviews by phone, classical face-to-face interviews, online surveys, or a mix thereof. Each design has some benefits, but also unique drawbacks that have to be accounted for.

Election polling is shaped by diverse sources of error that propagate to subsequent analyses. In this context, it is important to differentiate between nowcasting (i.e. the description of the *current* electoral mood) and forecasting (i.e. the prediction of the next election result). Nowcasting is affected by errors in sampling design and/or individuals' response behavior. (Sampling-based) Forecasting underlies the same problems, with additional uncertainty about how the current situation evolves until election day.

## 2.1 Nowcasting Errors

The issues with nowcasting are similar to the issues of survey sampling. In the concept of total survey error (see e.g. Biemer 2010) a distinction is made between sampling error and non-sampling error. When uncertainties are specified, it is common practice to provide ranges based on sampling error only. However, this neglects non-sampling errors, which in particular include non-specification errors, non-response, coverage errors, and measurement errors. Non-sampling errors are structurally related to the survey type and method. For example, in a telephone survey, where only landlines are contacted, people without a landline cannot be reached. If this technical accessibility of the interviewees is related to their voting decision, such problems inevitably lead to biases.

There is extensive literature on correcting biases caused by non-response and the lack of coverage. The most common methods are post-stratification and weighting methods. In the case of nowcasting, it is common practice (in Germany) to report weighted results instead of raw values (see wahlrecht.de). The main assumption behind such procedures is that sampling biases in the variables of interest can be attributed to biases in the characteristics used for weighting. If important characteristics were not recorded in the survey, biases can only be corrected inadequately. Moreover, the relationship among the weighting variables must be properly taken into account, including possible interactions. Another limitation is the potential lack of necessary information about the weighting variables in the population.

In addition to survey design problems, other biases may arise from response behavior since most nowcasting surveys are interviews in which participants may – consciously or unconsciously – make false statements or refuse to answer. The implication for data quality is difficult to assess since it relies on knowledge about the relationship between participation in the survey and the characteristic of interest. Undecided voters pose another problem, especially if the undecidedness is associated with party preference, leading to an underrepresentation of voters with specific political preferences. While weighting might be a possible way to cope with such issues, subsequent conceptual problems may again lead to other errors, if correction is inadequate.

Given the inherent limitations of election polling, it is apparent that every survey is affected by one or more of these problems. Assessing their magnitude is difficult since the 'truth' is unknown at the time of the nowcast.

## 2.2 Forecasting Errors

Forecasting of election results is performed with different methods. Established methods include model-based forecasts that take economic developments or expert knowledge into account. Instead, we focus on survey-based forecasting, which can be seen as an extrapolation of a nowcast into the future, i.e. for election day. While all previously mentioned problems with nowcasting also apply to forecasts (with some restrictions), further problems arise from attempting to predict the future. In contrast to nowcasts, forecast strategies can be evaluated using previous elections.

The main issue of forecasting is the transferability of results from the present to the future. Individual polls may be shaped by events that affect the political mood in the short term only. Further, it is impossible to account for potential future events which may (strongly) influence electoral behavior. A prominent example of the latter is the letter from James Comey (Silver, 2017), published shortly before the 2016 U.S. presidential election.

Most forecasters had already made their call and accordingly could not include the potential impact of the letter in their predictions.

While describing potential errors in a forecast is possible, quantifying the resulting uncertainty remains a major problem. Methods that account for uncertainty about future events are an area of active research.

## 3. Presentation of Nowcasts and Forecasts

The standard way of reporting results of opinion polls in (German) media is shown in Figure 1 (left pane), where a bar chart depicts the reported party shares of one poll. This way of visualization inherently ignores uncertainty. Moreover, the underlying uncertainty is mostly communicated in the accompanying text only, separated from the visual presentation. This exacerbates its proper perception. Such reports are then often accompanied by oversimplifying headlines like "No majority for the grand coalition" (FAZ, 2018, own translation), completely disregarding uncertainty. Especially complex issues like seat majorities in parliament are often inadequately visualized in the media. For example, the visualization in the right pane of Figure 1 neglects relevant characteristics of the German election system (i.e. parties with less than 5% voter share are not represented in parliament).

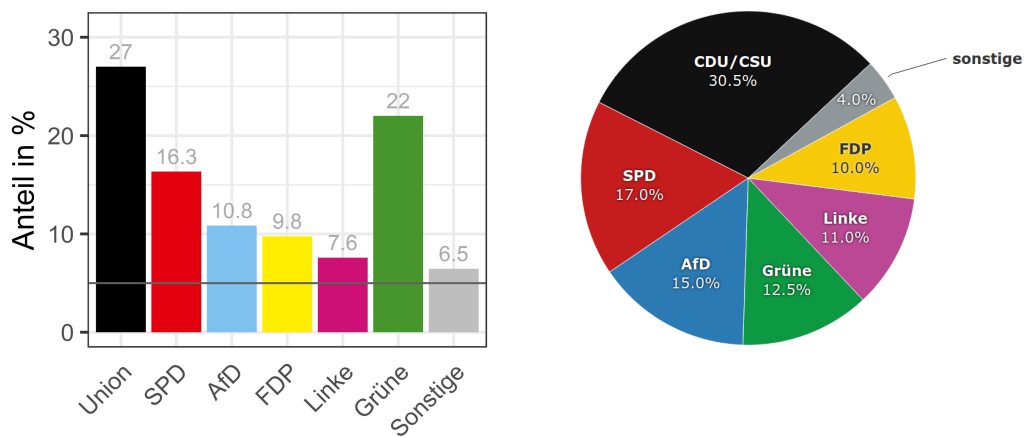

**Figure 1** Schematic bar plot of reported party shares (left pane) and FAZ visualization for whether Union and SPD jointly reach a majority in parliament (right; FAZ, 2018).

Also reports on changes in the current nowcast are often problematic. Oftentimes, only very minor differences in party shares – clearly within the expected uncertainty – are overinterpreted. Interpretations like the party "[...] SPD fell by half a percentage point to 17 percent" (FAZ, 2018) without simultaneously highlighting uncertainty are strongly misleading. Moreover, such developments over time are rarely communicated visually (i.e. time-series visualization).

Additionally, media reports often fail to differentiate between nowcasting and forecasting, using similar verbalizations for both. Especially the aforementioned additional sources of uncertainty or biases regarding forecasts are rarely pointed out.

A general problem for the media is that the average voter is not well educated regarding the assessment of uncertainties. Graefe (2021) provides empirical evidence indicating that the average German voter struggles to derive the desired information (e.g. what their strategic vote should be) from election polls in their currently communicated form. Providing the sampling error did not improve the participants' assessment. Furthermore, the participants generally overestimated a party's chance to be represented in parliament when it polled below the 5% threshold.

## 4. State-of-the-Art Concepts for Visual Presentation

In the following, we present several conceptual ideas for alternative communication of surveys. The central difficulty is to create visualizations that explicitly incorporate uncertainty, but which can also be understood easily by a layman.

### 4.1 Probability of Event and Uncertainties

One approach for improving the presentation of survey results is to shift the focus away from (projected) party shares. In many multi-party systems, single-party governments are improbable and coalitions are more of interest. Also, specific regulations are common that complicate any direct evaluation. For example, the 5% threshold in Germany adds uncertainty about whether a party is even represented in parliament. Useful alternative metrics are so-called Probabilities of Events (PoE), which can, for example, be estimated using Bayesian methods (see Bender et al., 2018; Bauer et al., 2020). These indicate the probability of a certain event of interest, e.g. that, based on a current poll and its underlying uncertainty, a specific coalition would obtain a majority. While standard reporting of election polls ignores uncertainty and the complex relationships that lead to a majority (e.g. the amount of redistributed votes; the closeness to the 5% threshold), PoEs contain information both about the likelihood and the uncertainty of an event and condense this information into single numbers.

Figure 2 shows an application of PoEs, each bar representing the probability for one specific coalition reaching a majority. The color of each bar highlights the party with the highest share. Outcomes in which a subset of the parties suffices to reach a majority (highlighted in light gray) are differentiated from such where the whole coalition is needed (full color).

Figure 3 presents the uncertainty in a simplified manner, by showing the distribution of expected (joint) seats for selected parties in parliament. Instead of PoEs, the concrete seat shares are shown for each event of interest. This properly visualizes the uncertainty in more complex situations, e.g. when one of the considered parties is at risk of not passing the threshold. The area representing the occurrence of the event can then be color-coded, which allows the simultaneous representation of PoEs.

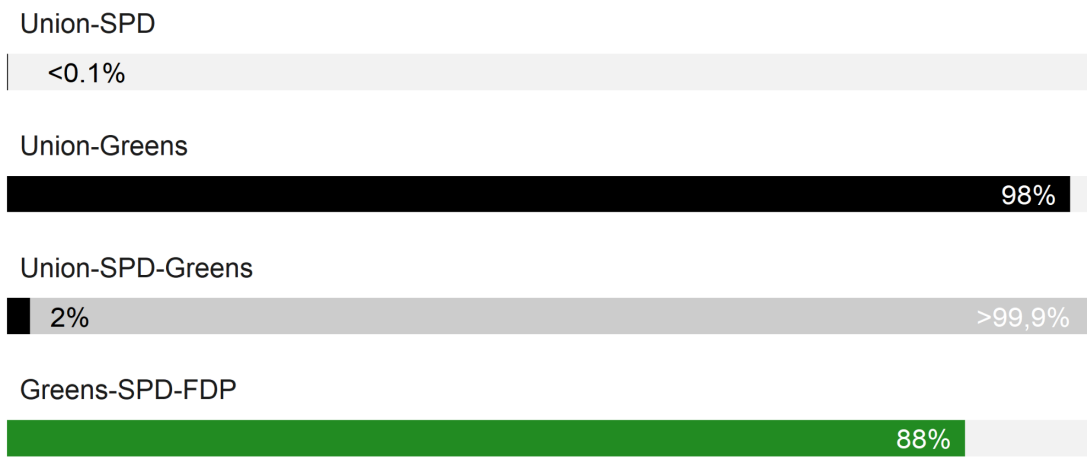

**Figure 2** Probabilities of Events (PoEs; nowcast) for selected coalitions in Germany based on one survey. The colors represent the strongest party within each coalition; the probability of a subset of parties achieving a majority is represented in light gray.

The most natural way of visualizing uncertainty when communicating surveys is to show multiple possible outcomes. Figure 4 shows six realizations of seat shares in German parliament, sampled based on one survey. The parties are arranged so that the coalition of interest and whether it reached a majority can easily be inferred. The larger the number of realizations, the easier it is to grasp the overall uncertainty. Accordingly, a trade-off must be made between accessibility and adequate uncertainty mapping. The concept behind Figure 4 can also be used for interactive presentations.

An extension of Figure 3 are so-called ridgeline plots (Bauer et al., 2020), which visualize the surveys' uncertainties and the development over time simultaneously (see Figure 5, left pane). For each survey, the distribution of expected seats is shown. The y-axis represents time, with more recent polls positioned closer to the bottom. In addition to visualizing sampling uncertainty, ridgeline plots also depict the variation over time with its implied uncertainty.

The ideas presented above are a more adequate representation of uncertainty than the usual focus on party shares. Corresponding ideas are not new and were used, for example, by FiveThirtyEight in the context of the 2016 and 2020 U.S. elections. As a general recommendation, we advocate the use of multiple types of visualizations to use their different focal points as an active element of communication. For example, we use the right pane of Figure 5 to highlight the PoE over time as one key aspect of the ridgeline plot.

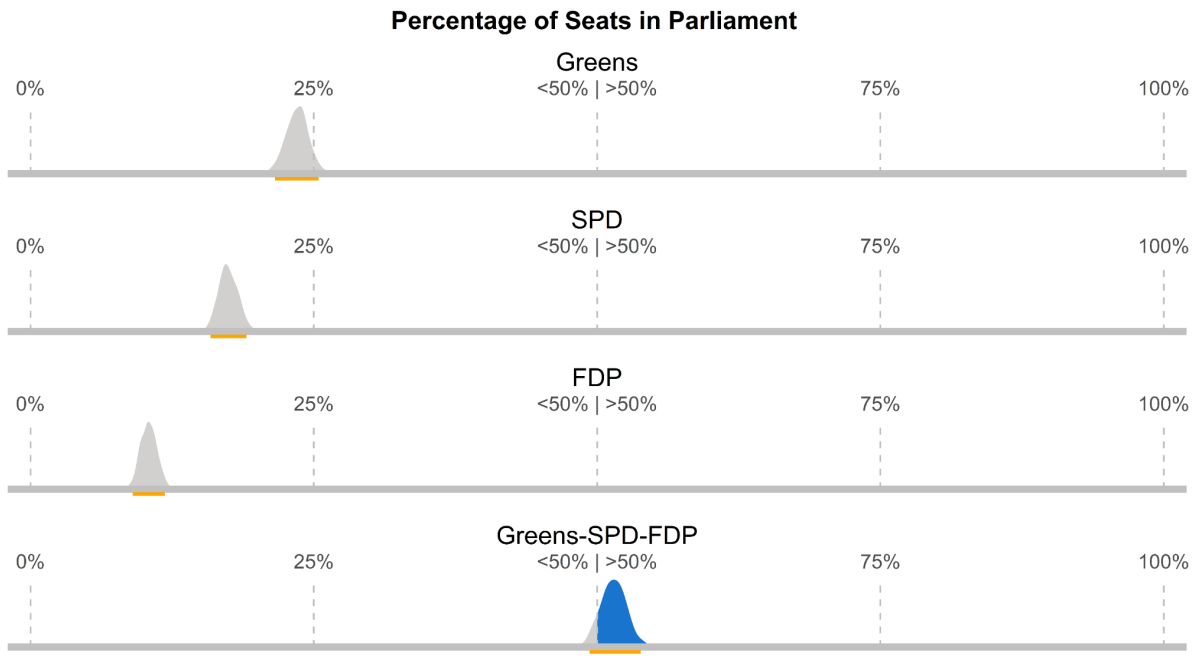

**Figure 3** Distribution of the expected seat shares (nowcast) for selected coalitions in Germany. The blue areas of the distributions represent a reached majority, the orange bar the 95% confidence region.

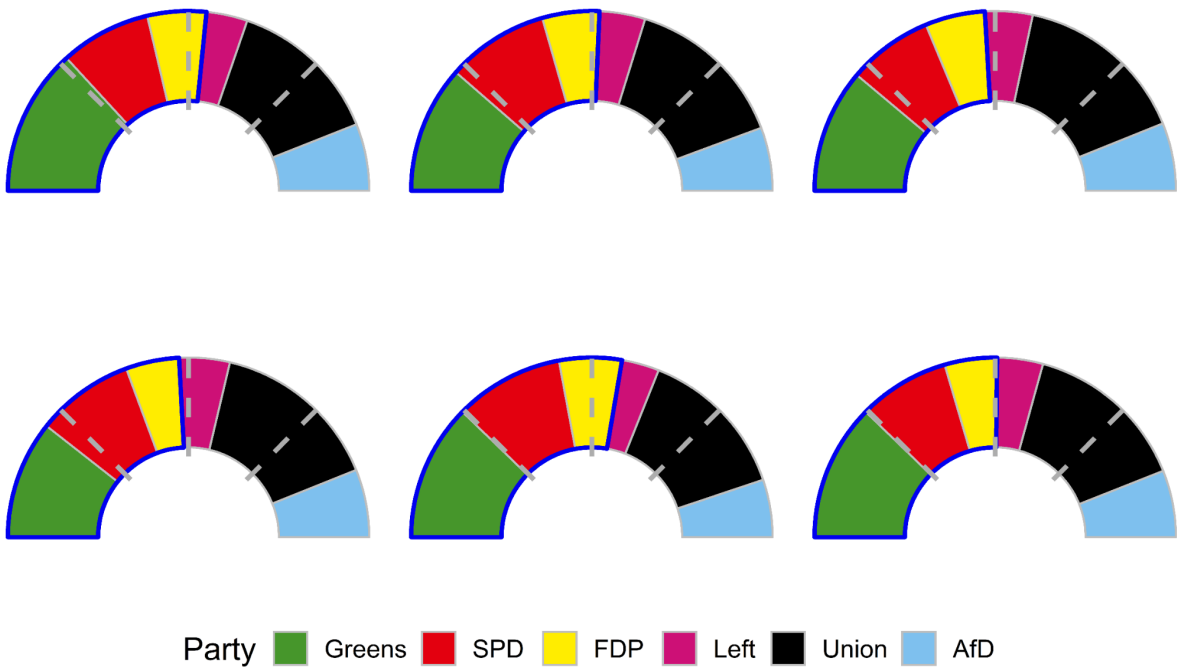

**Figure 4** Nowcast-based potential parliaments in Germany. Parties are color-coded and grouped to ease the assessment of a joint Greens-SPD-FDP majority. The gray dashed lines represent the quartiles.

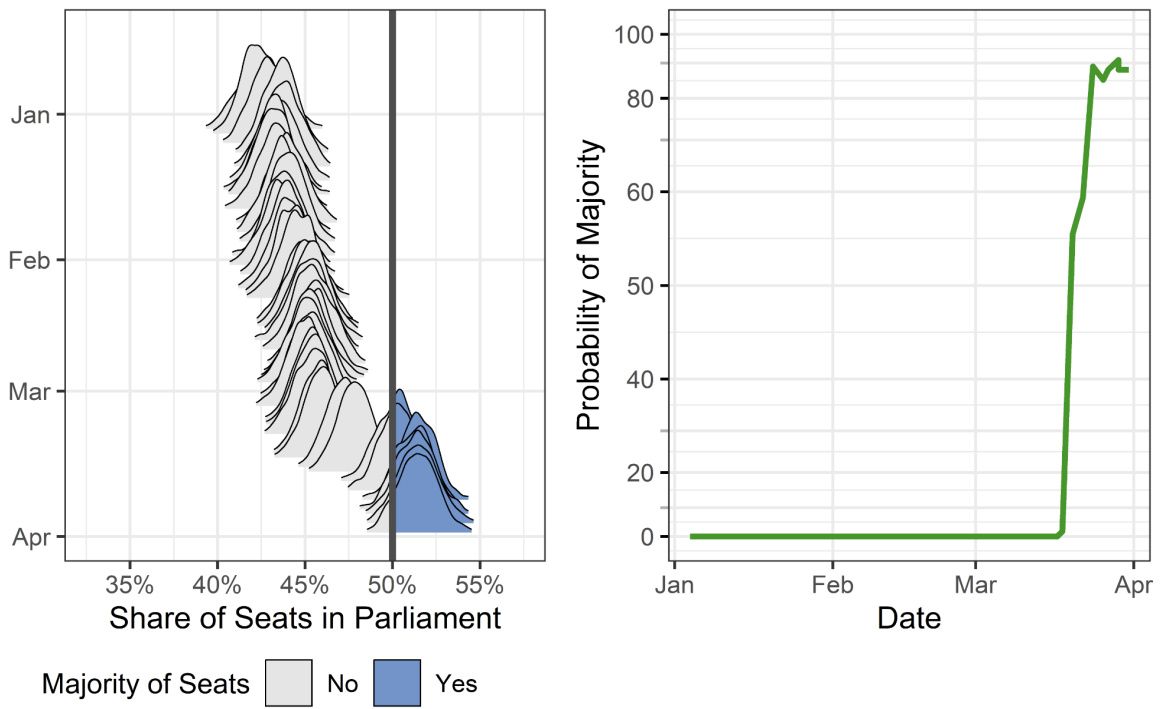

**Figure 5** Nowcast of expected seat shares for a Green-SPD-FDP coalition (ridgeline plot, left pane) and PoE over time for the coalition's majority (right, with a nonlinear y-axis). Blue areas in the left pane mark reached majorities, the solid black line 50%.

4.2 Forecast Visualizations

The different focus of forecasts requires further adjustments to visualize the additional uncertainty. A possible extension of the classic survey presentation with a forecast model is shown in Figure 6. The graph is similar to the usual survey presentation but includes the surveys' uncertainty in the form of an uncertainty area. The visualization also highlights the gap between the current day and Election Day. The increasing uncertainty of the outcome is visualized by a broadening uncertainty area. A comparable representation with a focus on PoE, avoiding the limitation of focusing on the individual party shares, is easily possible.

Another possibility for visualizing forecasts is schematically shown in Figure 7. It is based on the previously presented ridgeline plot. The left pane depicts the nowcasts already shown in Figure 5, the right pane shows the forecasts based on the respective nowcasts. Here, larger uncertainty is expressed by a wider distribution, and the uncertainty increases with greater distance to Election Day. The direct comparison of now- and forecast then allows the visual communication of the increased uncertainty.

However, both approaches share a fundamental problem: the assumption that uncertainty can be adequately estimated. Especially unknown future events are a problem in determining this uncertainty. But despite this limitation, both approaches are generally suitable to communicate an increase in uncertainty.

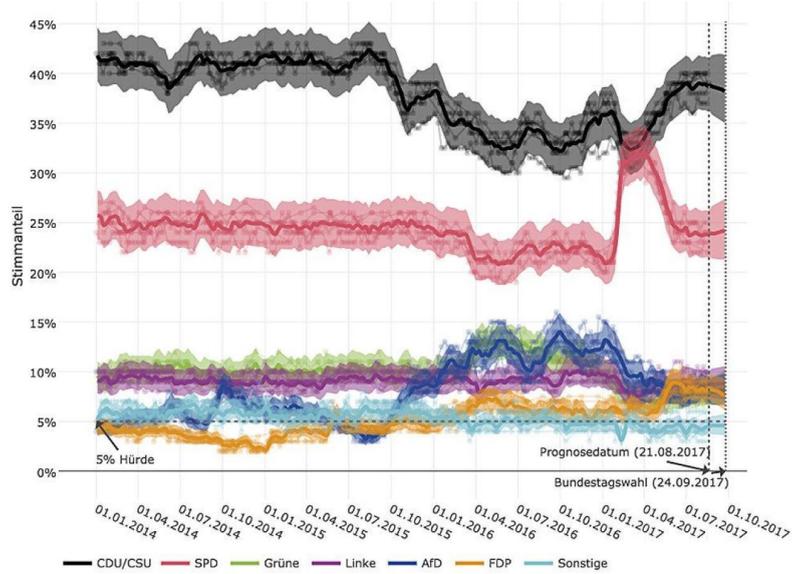

**Figure 6** INWT Statistics forecast for the German federal election 2017 (date of forecast 21.08.2017; Hendrich, 2017). The current day is marked by a vertical line, and Election Day represents the end of the x-axis. Each party is color-coded, confidence regions are depicted by colored areas, the dots represent individual surveys.

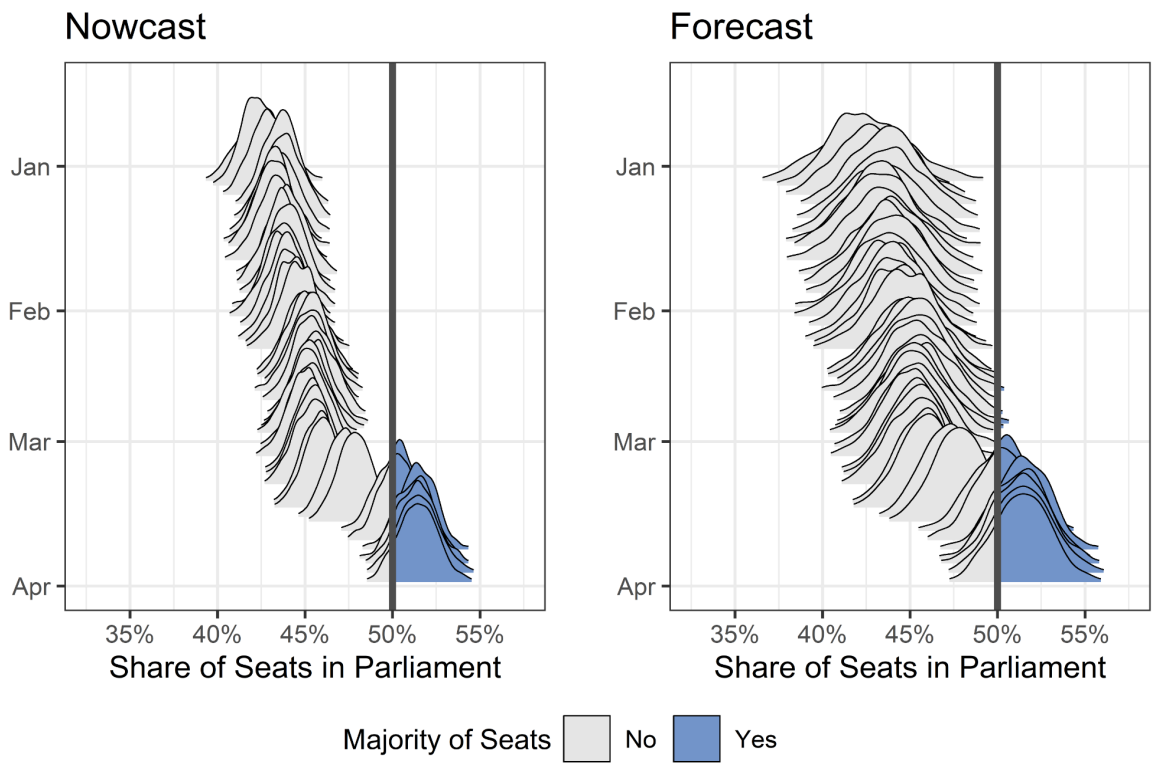

**Figure 7** Nowcast (left pane) and forecast (right) for a majority of a Green-SPD-FDP coalition in Germany using ridgeline plots. The blue areas in the left pane mark reached majorities, the solid black line 50%. Note that this is a schematic representation; the forecast is not based on an actual forecast model.

## 5. Discussion

To summarize, many established ways of presenting opinion polls often only inadequately reflect uncertainty. Uncertainty is, if at all, explained in a rather technical way and is often separated from the actual communicated content. While some providers in the U.S. are actively working on this communication problem, no such effort is observed in Germany. This is aggravated by the fact that the public's understanding of probabilities is still insufficient.

In our opinion, the presentation of uncertainty too often focuses on sampling uncertainty alone. An election survey, whether for now- or forecasting, is subject to multiple sources of uncertainty and sampling uncertainty only inadequately represents them. A discussion of the sources of uncertainty and their implications is imperative, as only then can they be properly considered in visual communication.

One way to improve communication is to shift the focus away from individual parties to relevant events. PoEs provide a way to represent complex issues in one metric, with associated benefits in visual communication. However, because of the general public's struggles with understanding probabilities, PoEs should not be used exclusively, but in conjunction with other representations. Our suggestions are densities of seat distributions and ridgeline plots based on them. These graphical concepts are also usable when PoEs further account for non-sampling errors. Together, they allow for a better communication of uncertainties without being overly complex individually.

The established ways of communicating election surveys mostly do not do justice to the problem. While further research is needed on the perception of the discussed graphics and the associated probabilities to find ideal ways of communication, the difficulties observed in the current understanding of (event) probabilities cannot be an argument against their use.